# On Outage Probability and Diversity-Multiplexing Tradeoff in MIMO Relay Channels

Sergey Loyka and Georgy Levin

*Abstract*—Fading MIMO relay channels are studied analytically, when the source and destination are equipped with multiple antennas and the relays have a single one. Compact closed-form expressions are obtained for the outage probability under i.i.d. and correlated Rayleigh-fading links. Low-outage approximations are derived, which reveal a number of insights, including the impact of correlation, of the number of antennas, of relay noise and of relaying protocol. The effect of correlation is shown to be negligible, unless the channel becomes almost fully correlated. The SNR loss of relay fading channels compared to the AWGN channel is quantified. The SNR-asymptotic diversity-multiplexing tradeoff (DMT) is obtained for a broad class of fading distributions, including, as special cases, Rayleigh, Rice, Nakagami, Weibull, which may be non-identical, spatially correlated and/or non-zero mean. The DMT is shown to depend not on a particular fading distribution, but rather on its polynomial behavior near zero, and is the same for the simple "amplify-and-forward" protocol and more complicated "decode-and-forward" one with capacity achieving codes, i.e. the full processing capability at the relay does not help to improve the DMT. There is however a significant difference between the SNR-asymptotic DMT and the finite-SNR outage performance: while the former is not improved by using an extra antenna on either side, the latter can be significantly improved and, in particular, an extra antenna can be traded-off for a full processing capability at the relay. The results are extended to the multi-relay channels with selection relaying and typical outage events are identified.

*Index Terms*—MIMO, relay channel, outage probability, channel capacity, diversity-multiplexing tradeoff.

## I. INTRODUCTION

COOPERATIVE communication strategies have recently emerged as an efficient way to exploit multi-user diversity available in wireless networks to significantly improve their performance [1]-[4]. Starting from the pioneering work in [1], a number of efficient protocols, distributed space-time codes and signal processing strategies have been proposed [5]-[16]. While the research was initially concentrated on the single-antenna systems [1]-[3][5][12]-[14], the emphasis has recently shifted towards multi-antenna systems [6]-[11],[15]. Typical performance metrics in a fading channel include the outage capacity or the outage probability [1][11]-[14], mean (ergodic) capacity [9] and error rates when specific codes/modulation formats are studied [3][10][15]. Due to the complexity of the analysis, the performance studies have been mainly limited to independent (but not necessarily identically distributed) Rayleigh-fading channels.

Since the MIMO systems present an additional level of difficulty in terms of performance evaluation for all the metrics, an elegant framework termed "diversity-multiplexing tradeoff" (DMT) has been proposed in [17], which allows one to quantify the system performance in terms of two principal gains, diversity and multiplexing, available in a slow-fading MIMO channel when $SNR \rightarrow \infty$ [18]. Many systems, for which the outage probability/capacity analysis is illusive, can be characterized and compared via their respective DMTs. This framework has been successfully applied to relay channels as well [5]-[8][11]. Since the amplify-and-forward (AF) and decode-and-forward (DF) protocols proposed in [1] are not DMT-optimal at high multiplexing gains, a number of new protocols were proposed in [5] and their DMT has been investigated. This work was further extended to the multi-antenna terminals in [6]-[8][11][34]. While most of the DMT-based studies are limited to the $SNR \rightarrow \infty$ regime, the finite-SNR DMT of the relay channel with independent, Rayleigh-fading links and single-antenna terminals have been studied in [28].

The outage capacity of relay channels with independent Rayleigh fading links has been studied in [13] in the low-SNR regime, and it was shown that the AF and DF protocols are sub-optimal, and a new strategy, the bursty amplify-and-forward (BAF), was proposed and shown to be optimal, to the $1^{st}$ order, in the low-SNR low-outage regime, which also includes the case of multiple relays. These results have been extended to the case of relay channels with generic fading distribution (but the links are still required to be independent) admitting low-outage Taylor expansion in [14], where it was shown that the BAF still achieves the outage capacity to the $1^{st}$ order (but not to the second order) in the low-SNR low-outage regime. The case of selection relaying (where the best relay only is used) has also been studied and the BAF strategy has been shown to be optimal in this case as well. Both of these works consider single-antenna terminals.

In all the studies above, a Rayleigh or Rician-fading channel model with independent links has been employed. The only exception is [14], where the outage probability/capacity has been studied for a generic fading distribution but the analysis was limited to the low-SNR regime and the links are still required to be independent.

In this paper, we consider the source and the destination of the relay channel equipped with multiple antennas and relay nodes equipped with a single antenna (e.g. due to complexity constraints). We allow the fading to be non-identical, correlated Rayleigh or of generic distribution (non-Rayleigh/Rice), and consider amplify-and-forward and decode-and-forward protocols. The contributions of the paper are as follows:

• The outage probability and outage capacity of correlated and/or non-identical Rayleigh fading are obtained in a closed form (Theorems 1 and 2 in Section III), including insightful low-outage approximations (Corollaries 1 and 2). This allows one to draw important design guidelines and also to establish the limitations of SNR-asymptotic DMT-based designs: two systems with identical DMT may have vastly different outage performance, e.g. while the DMT of the $1 \times 1$, $2 \times 1$ and $1 \times 2$ channels is the same, an additional antenna results in significantly lower outage probability (in fact, the ratio of $1 \times 1$ and $1 \times 2$ channel outage probabilities grows unbounded as SNR $\to \infty$). This is equivalent to a significant SNR gain (about 10 dB at the outage probability = $10^{-3}$), not captured by the DMT framework. An additional antenna can also be traded off for the full processing capability at the relay. Furthermore, an extra transmit rather than receive antenna is preferable, since, unlike the latter, the former eliminates the negative effect of relay noise in the low outage regime. Therefore, one concludes that the DMT framework is ill-suited for some relay channels. A study of the effect of correlation demonstrates that, unlike full-rank MIMO channels, it has here a negligible impact on the performance unless the channel is nearly-fully correlated. Under certain conditions, the relay channel is shown to be equivalent to the maximum ratio combining one with an extra array gain.

• The SNR-asymptotic DMT of the relay channels in the AF and DF modes is obtained for a broad class of fading distributions including, as special cases, Rayleigh, Ricean, Nakagami, and Weibull, which can be non-identical, non-zero mean and/or spatially correlated (see Theorems 3-5 in Section IV). The AF and DF systems have the same DMT, which depends on the minimum diversity order only. This generalizes/extends the known DMT or low-SNR results mentioned above.

• These results are extended to the multi-relay channels with selection relaying. Typical outage events in relay channels are identified.

The rest of the paper is organized as follows. Section II introduces the basic system model. Outage probability and outage capacity are studied analytically in Section III. The SNR-asymptotic DMT is considered in Section IV for a broad class of fading distributions in the AF and DF modes and its limitations are pointed out. Multi-relay channels with selection relaying are considered in Section V. Finally, Section VI concludes the paper and the appendix details derivations.

## II. SYSTEM MODEL

We consider a MIMO relay channel where the source (transmitter) and the destination (receiver) are equipped with multiple antennas and a relay node equipped with a single antenna (see Fig. 1). While both the amplify-and-forward and decode-and-forward protocols are considered, the former will be assumed for simplicity of exposition, unless indicated otherwise, with a fixed gain relay (this is motivated by the fact that it is simpler to implement). We assume no direct source-destination link. This is motivated by the fact that the direct link may be much weaker than the relay one (e.g. no line-of-sight etc.) and thus can be neglected (this is the case when the relay link is needed most) [3]; it also makes the analysis tractable. The case of multiple relay nodes is considered in Section V.

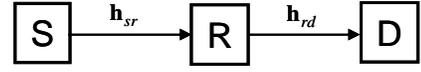

Fig. 1. Relay channel with a single relay node (single-antenna) and multiple-antenna source (transmitter) and destination (receiver).

The standard baseband system model of a frequency-flat block-fading (quasi-static) relay channel in the amplify-and-forward mode is given by,

$$\mathbf{y} = \sqrt{K_r G_{rd} G_{sr}} \mathbf{h}_{rd} \mathbf{h}_{sr}^+ \mathbf{x} + \sqrt{K_r G_{rd}} \mathbf{h}_{rd} \xi_r + \boldsymbol{\xi} \quad (1)$$

where $\mathbf{x}$ and $\mathbf{y}$ are the source (transmitter) and destination (receiver) symbol vectors, $\mathbf{h}_{sr}^+$ and $\mathbf{h}_{rd}$ are the source-relay and relay-destination normalized channels (i.e. they include the multipath fading but not the average path loss), $^+$ denotes Hermitian conjugation, $G_{sr}$ and $G_{rd}$ are the source-relay and relay-destination average path loss factors, $K_r$ is a fixed relay gain, $\xi_r \sim \mathcal{CN}(0, \sigma_r^2)$ and $\boldsymbol{\xi} \sim \mathcal{CN}(\mathbf{0}, \sigma_0^2 \mathbf{I})$ are relay and destination AWGN noise of variance $\sigma_r^2$ and $\sigma_0^2$ respectively, and independent of each other. While analyzing the outage probability, we assume that $\mathbf{h}_{sr}^+$ and $\mathbf{h}_{rd}$ are i.i.d. or correlated Rayleigh-fading and are independent of each other, and while analyzing the SNR-asymptotic DMT, $\mathbf{h}_{sr}^+$ and $\mathbf{h}_{rd}$ are also assumed to be independent of each other but with a generic fading distribution of a polynomial near-zero behavior (most known models are in this class); $m$ and $n$ will denote the number of source and destination antennas. Note that the first term in (1) represents the signal received at the destination; $2^{nd}$ and $3^{rd}$ terms represent the relay noise propagated to the destination and the destination noise. The sufficient statistics for $\mathbf{y}$ is (see e.g. [18]),

$$z = \frac{\mathbf{h}_{rd}^+}{|\mathbf{h}_{rd}|} \mathbf{y} \quad (2)$$

$$= \sqrt{K_r G_{rd} G_{sr}} |\mathbf{h}_{rd}| \mathbf{h}_{sr}^+ \mathbf{x} + \sqrt{K_r G_{rd}} |\mathbf{h}_{rd}| \xi_r + \frac{\mathbf{h}_{rd}^+}{|\mathbf{h}_{rd}|} \boldsymbol{\xi}$$

where $|\mathbf{h}|^2 = \mathbf{h}^+ \mathbf{h}$, and the instantaneous SNR at the destination can be expressed as

$$\gamma' = \frac{K_r G_{rd} G_{sd} |\mathbf{h}_{rd}|^2 \mathbf{h}_{sr}^+ \mathbf{R}_x \mathbf{h}_{sr}}{\sigma_0^2 + K_r G_{rd} |\mathbf{h}_{rd}|^2 \sigma_r^2} \leq \frac{|\mathbf{h}_{rd}|^2 |\mathbf{h}_{sr}|^2}{1 + \alpha |\mathbf{h}_{rd}|^2} \gamma \quad (3)$$

where $\mathbf{R}_x = \overline{\mathbf{x}\mathbf{x}^+}$ is the covariance matrix of the transmitted signal, $\alpha = K_r G_{rd} \sigma_r^2 / \sigma_0^2$ is the ratio of the average relay noise propagated to the destination to the destination noise, and $\gamma = K_r G_{rd} G_{sr} \sigma_x^2 / \sigma_0^2$ is the average SNR at the destination, $\sigma_x^2 = tr\mathbf{R}_x = \overline{\mathbf{x}^+ \mathbf{x}}$ is the total transmitted power (at the source). The inequality in (3) follows from $\mathbf{h}_{sr}^+ \mathbf{R}_x \mathbf{h}_{sr} \leq |\mathbf{h}_{sr}|^2 \sigma_x^2$, and the equality is achieved when $\mathbf{R}_x = \sigma_x^2 \mathbf{h}_{sr} \mathbf{h}_{sr}^+ / |\mathbf{h}_{sr}|^2$, i.e. beamforming from the source to the relay, $\mathbf{x} = s \mathbf{h}_{sr} / |\mathbf{h}_{sr}|$, where $s$ is the scalar transmitted symbol of the total power $\sigma_x^2$. This, of course, requires channel state information (CSI) at the source. When no such information is available, a sensible transmission strategy is isotropic [18], i.e. $\mathbf{R}_x = \sigma_x^2 \mathbf{I}/m$. In this case, the SNR is

$$\gamma' = \frac{|\mathbf{h}_{rd}|^2 |\mathbf{h}_{sr}|^2}{1 + \alpha |\mathbf{h}_{rd}|^2} \frac{\gamma}{m} \quad (4)$$

Comparing this to the upper bound in (3), one concludes that the source CSI brings in an m-fold SNR gain, but does not change the statistics of the instantaneous SNR otherwise and, therefore, the outage probability or outage capacity differ by a constant SNR shift, and the diversity-multiplexing tradeoff is the same in both cases. Below we assume the source CSI so that the instantaneous SNR is

$$\gamma' = \frac{|\mathbf{h}_{rd}|^2 |\mathbf{h}_{sr}|^2}{1 + \alpha |\mathbf{h}_{rd}|^2} \gamma \quad (5)$$

The instantaneous channel capacity (in nats/s/Hz) can now be expressed as[1] $C = \ln(1 + \gamma')$ and the channel outage probability [2], i.e. the probability that the channel is not able to support a target rate $R$, is

$$P_{out} = \Pr\{C < R\} = \Pr\left\{ \frac{|\mathbf{h}_{rd}|^2 |\mathbf{h}_{sr}|^2}{1 + \alpha |\mathbf{h}_{rd}|^2} < \frac{e^R - 1}{\gamma} \right\} \quad (6)$$

It follows from (3), (5), (6) that the optimum (i.e. capacity-achieving) transmission strategy is the beamforming at the source (towards the relay) and the maximum ratio combining at the destination.

## III. Outage Probability, Capacity and DMT at Finite SNR

Since the finite-SNR analysis is not feasible for a generic fading distribution, in this section we consider Rayleigh-fading links (e.g. source-relay and relay-destination), which may be non-identical and/or correlated. First, the outage probability is investigated for fixed $R$. Then, the finite-SNR diversity-multiplexing tradeoff is discussed.

### A. Outage probability

*Theorem 1:* Let $\mathbf{h}_{rd}$ and $\mathbf{h}_{sr}$ be mutually independent circular-symmetric correlated Gaussian random vectors. When the eigenvalues of the source-relay and relay-destination link correlation matrices $\mathbf{R}_{sr(rd)} = \overline{\mathbf{h}_{sr(rd)} \mathbf{h}_{sr(rd)}^+}$ are non-zero and distinct, the outage probability of the single-relay, correlated channel in the amplify-and-forward mode is

$$P_{out} = 1 - \sum_{i=1}^{m}\sum_{j=1}^{n} A_i B_j e^{-\alpha x/\lambda_j} \sqrt{\frac{4x}{\lambda_i \eta_j}} K_1\left(\sqrt{\frac{4x}{\lambda_i \eta_j}}\right) \quad (7)$$

where $K_N(x)$ is the $N$-th order modified Bessel function of the second kind [27], $A_i$ and $B_j$ are the coefficients of a partial fraction decomposition given by $A_k = \prod_{i \neq k} \lambda_k/(\lambda_k - \lambda_i)$ and likewise for $B$, $\lambda_i$ and $\eta_j$ are the eigenvalues of $\mathbf{R}_{sr(rd)}$, and $x = (e^R - 1)/\gamma$.

---

[1] Note that the channels in (1) and (2) have the same capacity (since z is a sufficient statistics).

[2] It is also the best achievable codeword error probability [35][36], i.e. a fundamental performance limit. Realistic codes can be handled via the SNR gap to capacity [32].

*Proof:* see Appendix. ∎

While this expression holds when eigenvalues are different and non-zero, the case of identical eigenvalues can be handled by the limiting transition using L'Hopital's rule (since the outage probability is a continuous function of eigenvalues), and zero eigenvalues should be simply excluded, i.e. $m, n$ should be treated as the ranks of corresponding correlation matrices.

To get some insight, let us now consider the low-outage regime.

*Corollary 1.1:* The behavior of $P_{out}$ in (7) in the low-outage regime $x \to 0$[3] is determined by the $\min(m,n)$ term and is given by:

$$\begin{aligned}
P_{out} &= a_1 x^m + o(x^m), \ m < n \quad (8) \\
&= a_2 x^n + o(x^n), \ m > n \\
&= \left(a_3 + b_3 \ln \frac{1}{x}\right) x^m + o(x^m), \ m = n
\end{aligned}$$

where $a_1$, $a_2$, $a_3$ and $b_3$ are constants independent of $x$,

$$a_1 = \frac{\alpha^m}{m! \det \mathbf{R}_{sr}} + \frac{(-1)^{m+1}}{\det \mathbf{R}_{sr}} \sum_{k=1}^{m}\sum_{i=1}^{n} \frac{B_i \ln \eta_i}{\eta_i^k} D_{km}(\alpha),$$

$$a_2 = \frac{(-1)^{n+1}}{n!(n-1)! \det \mathbf{R}_{rd}} \sum_{j=1}^{m} \frac{A_j \ln \lambda_j}{\lambda_j^n},$$

$$a_3 = \frac{\alpha^m}{m! \det \mathbf{R}_{sr}} + \frac{(-1)^{m+1}}{m!(m-1)! \det \mathbf{R}_{rd}} \sum_{j=1}^{m} \frac{A_j}{\lambda_j^m} \ln \lambda_j$$

$$+ b_3 \Psi_m + \frac{(-1)^{m+1}}{\det \mathbf{R}_{sr}} \sum_{k=1}^{m}\sum_{i=1}^{m} \frac{B_i \ln \eta_i}{\eta_i^k} D_{km}(\alpha),$$

$$b_3 = \frac{1}{m!(m-1)!} \frac{1}{\det \mathbf{R}_{sr} \det \mathbf{R}_{rd}},$$

$$D_{kl}(\alpha) = \frac{(-1)^{l-k} \alpha^{l-k}}{(l-k)!(k-1)!k!},$$

and $\Psi_k = \psi(k) + \psi(k+1)$, $\psi(1) = -\mathcal{C}$, $\psi(k) = -\mathcal{C} + \sum_{i=1}^{k-1} \frac{1}{i}$ for $k \geq 2$, and $\mathcal{C} \approx 0.577$ is Euler's constant.

*Proof:* see Appendix. ∎

It follows from (8) that the weakest link (with the lowest diversity order) dominates the outage performance, so that the channel diversity order is $\min(m,n)$. When $m = n$, both links contribute almost equally to the outage and the unusual term $\ln \frac{1}{x}$ emerges, which has a profound negative impact on the outage probability in the low outage regime ($x \ll 1$) and cannot be found in the regular (no relay) Rayleigh-fading channels. Note also that $P_{out}$ in (8) does not depend on $\alpha$ when $m > n$, i.e. extra Tx antenna(s) eliminate the negative effect of relay noise in the low-outage regime.

In a typical wireless system, the average path loss is about $50 \ldots 100$ dB or more (unless the transmitter and the receiver are very close to each other) [23], i.e. $G_{rd} \approx 10^{-5} \ldots 10^{-10}$ so that

$$\alpha = K_r G_{rd} \sigma_r^2 / \sigma_0^2 \ll 1 \quad (9)$$

---

[3] this requirement means that $\gamma \gg e^R - 1$, which is equivalent to $\gamma \to \infty$ under fixed $R$ (i.e. high average SNR) but also may hold at low SNR as well, when $R \ll \gamma \ll 1$.

when the relay gain $K_r$ and $\sigma_r^2/\sigma_0^2$ are not too large[4]. Motivated by this, we note that the channel in (1) and the outage probability in (7) are the same as those of the single-keyhole model [24], [25, Theorem 3.1] when $\alpha \to 0$, and the corresponding keyhole (double-scattering) channel based results also apply to the single-relay channel, e.g. the space-time codes [26] or diversity-multiplexing tradeoff [19][20]. Under the condition in (9), (7) can be expressed, after some straightforward but lengthy manipulations, as follows

$$P_{out} = \sum_{k=\min(m,n)}^{\infty} \frac{x^k}{k!(k-1)!} \sum_{i,j} \frac{A_j B_i}{(\lambda_j \eta_i)^k} \left( \ln \frac{\lambda_j \eta_i}{x} + \Psi_k \right) \quad (10)$$

In the low outage regime, $x \to 0$, the $\min(m,n)$ term dominates and (10) reduces to

$$P_{out} = \frac{x^m}{m!(m-1)!} \frac{\ln \frac{1}{x} + b_m}{\det \mathbf{R}_{rd} \det \mathbf{R}_{sr}} + o(x^m), \ m = n, \quad (11)$$

$$= \frac{x^m}{m!(m-1)!} \sum_{i=1}^{n} \frac{B_i \ln \eta_i}{\eta_i^m} \frac{(-1)^{m-1}}{\det \mathbf{R}_{sr}} + o(x^m), \ n > m$$

where $b_m = 1/m + 2\psi(1)$, and the $m > n$ case is obtained from the $n > m$ one via $m \leftrightarrow n$. Clearly, the diversity gain of the channel, at fixed rate $R$, is $d = \min(m,n)$. It is also clear that the outage probability increases with correlation (since $\det \mathbf{R}$ is maximum for uncorrelated channel and decreases with correlation) and the same conclusion holds for (8). It can be shown that the relay noise is negligible in (8) and the latter reduces to (11) when $\alpha \ll (m! \det \mathbf{R}_{sr})^{1/m}$ and either $\alpha \ll (m(m-1)\eta_i)^{-1}$ (for $m \geq 2$) or $\alpha \ll \sum_{i=1}^{n} B_i \ln \eta_i / \eta_i$ (for $m = 1$). Note that $(m! \det \mathbf{R}_{sr})^{1/m}$ decreases with correlation, so that a given relay noise may be negligible when the correlation is low, but not when it is high, which is another manifestation of the detrimental effect of the correlation. Fig. 2 shows the outage probability of a 2x2 correlated relay channel for different values of $\alpha$. Clearly, the effect of the relay noise is negligible unless $\alpha > 1$.

While the results above apply to a correlated channel with distinct eigenvalues and do not include the i.i.d. channel, the latter can be characterized as follows.

*Theorem 2:* Let $\mathbf{h}_{rd}$ and $\mathbf{h}_{sr}$ be i.i.d. mutually-independent circular-symmetric Gaussian random vectors. The outage probability of this single-relay channel in the amplify-and-forward mode is

$$P_{out} = 1 - \frac{2e^{-\alpha x}}{(n-1)!} \sum_{k=0}^{m-1} \sum_{i=0}^{k} \frac{\alpha^i x^{(k+i+n)/2}}{i!(k-i)!} K_{n+i-k}(\sqrt{4x}) \quad (12)$$

*Proof:* see Appendix. ∎

Note that (12) is easy to evaluate numerically since it contains finite sums and well-known special functions. It can be further expanded as a series in $x$:

*Corollary 2.1:* $P_{out}$ in Theorem 2 can be expressed as

$$P_{out} = \sum_{l=0}^{\infty} (f_l(\alpha) + g_l(\alpha) \ln x) \, x^{l+\min(m,n)}, \quad (13)$$

[4] As a practical example, $K_r = 60$ dB, $G_{rd} = -100$ dB, $\sigma_r^2 = \sigma_0^2 = -102$ dBm in a typical 3GPP UMTS scenario [33], so that $\alpha = 10^{-4} \ll 1$.

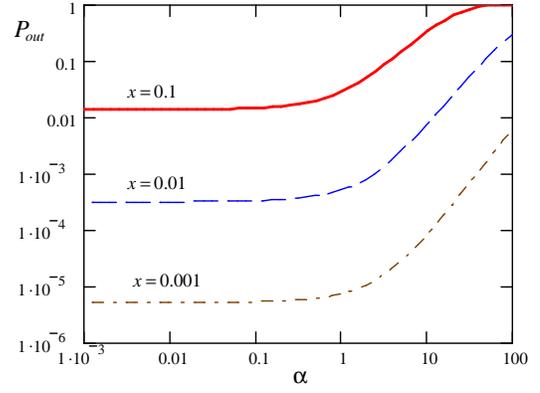

Fig. 2. Outage probability of the 2x2 channel vs. $\alpha$ for $x = 10^{-1}, 10^{-2}, 10^{-3}$; the normalized Tx/Rx correlation is $\rho = 0.5$. Note that $\alpha < 1$ has negligible impact on the outage probability and there is a significant increase in the $P_{out}$ when $\alpha > 1$, i.e. relay noise is important only when it is really high: $\sigma_r^2/\sigma_0^2 > 1/(K_r G_{rd})$. In a practically-important case of $\sigma_r^2 = \sigma_0^2$ [33], this reduces to $K_r G_{rd} > 1$ and the relay noise is negligible otherwise.

where $f_l(\alpha)$ and $g_l(\alpha)$ are independent of $x$ and are given in Appendix. The behavior of $P_{out}$ in the small outage region $x \to 0$ is determined by the $\min(m,n)$ term:

$$P_{out} = \frac{x^m}{(n-1)!} \sum_{k=0}^{m} \frac{\alpha^k (n-m+k-1)!}{(m-k)!k!} + o(x^m), \ m < n,$$

$$= \frac{x^n (m-n-1)!}{n!(m-1)!} + o(x^n), \ m > n, \quad (14)$$

$$= \frac{x^m}{(m-1)!} \left( \sum_{k=1}^{m} \frac{\alpha^k}{(m-k)!k} + \frac{\ln \frac{1}{x} + b_m}{m!} \right) + o(x^m),$$

$$m = n.$$

*Proof:* see Appendix. ∎

Clearly, the diversity order $d = \min(m,n)$ does not depend on $\alpha$ and the unusual term $\ln \frac{1}{x}$ is also present, which has a profound negative impact on the outage probability. $P_{out}$ does not depend on $\alpha$ when $m > n$, i.e. extra Tx antenna(s) eliminate the negative effect of relay noise in this case as well, and increases with $\alpha$ (i.e. with relay noise) otherwise.

When $\alpha \to 0$, $P_{out}$ becomes symmetrical with respect to $m$ and $n$, which is explained by the symmetry of the channel in (1) in this case, and (12) simplifies to

$$P_{out} = 1 - \frac{2}{(n-1)!} \sum_{k=0}^{m-1} \frac{x^{(k+n)/2}}{k!} K_{n-k}(\sqrt{4x}) \quad (15)$$

$$= \sum_{i=0}^{|n-m|-1} \mu_i x^{i+\min(m,n)} + \sum_{i=0}^{\infty} \beta_i x^{i+\max(m,n)} (\ln x - c_i)$$

where $\mu_i$, $\beta_i$ and $c_i$ are independent of $x$,

$$\mu_i = \frac{(-1)^i (|n-m|-i-1)!}{i!(\min(m,n)+i)(n-1)!(m-1)!}, \quad (16)$$

$$\beta_i = \frac{(-1)^{|n-m|+1}}{i!(\max(m,n)+i)(|n-m|+i)!(n-1)!(m-1)!}$$

$$c_i = \frac{1}{\max(m,n)+i} + \psi(i+1) + \psi(|n-m|+i+1)$$

and (14) reduces to[5]

$$P_{out} = \frac{\ln\frac{1}{x} + b_m}{m!(m-1)!} x^m + o(x^m), \quad m = n$$
$$= \frac{1}{m!}\left(\frac{x}{G}\right)^m + o(x^m), \quad n > m, \quad (17)$$

where $G = ((n-1)...(n-m))^{1/m} \geq 1$ and the $m > n$ case is obtained from the $n > m$ one via the substitution $m \leftrightarrow n$. Note that $P_{out} = P_{MRC}(x/G) \leq P_{MRC}(x)$ when $n > m$, where $P_{MRC}(x) \approx x^m/m!$ is the outage probability of $m$-branch maximum ratio combiner (MRC) in the i.i.d. Rayleigh-fading channel, so that the relay channel is better than ($n \geq 3$) or equal to ($n = 2$) the $m$-branch MRC channel in this case and, as far as the outage probability is concerned, the relay channel is equivalent to a cascade of a fading link (source-relay) and a non-fading relay-destination link with an SNR gain of beamforming at the destination equal to $G$. When $n \gg m$, $G \approx n$, i.e. a gain of n-element antenna array. Similar conclusions hold for the $n < m$ case via the substitution $m \leftrightarrow n$.

### B. Special Cases

To obtain some insight, let us now consider $1 \times 1$, $2 \times 1$ (2 Tx, 1 Rx antenna) and $1 \times 2$ i.i.d. channels. In these cases, (12) simplifies to

$$P_{out}^{1\times 1} = 1 - e^{-\alpha x}\sqrt{4x}K_1(\sqrt{4x}),$$
$$P_{out}^{2\times 1} = 1 - 2e^{-\alpha x}\left[(\sqrt{x} + \alpha x^{3/2})K_1(\sqrt{4x}) + xK_0(\sqrt{4x})\right]$$
$$P_{out}^{1\times 2} = 1 - 2xe^{-\alpha x}K_2(\sqrt{4x}). \quad (18)$$

At the low outage regime $\alpha x \ll 1$, this can be approximated as

$$P_{out}^{1\times 1} \approx x(\alpha + \ln\frac{1}{x}), \ P_{out}^{1\times 2} \approx x(1+\alpha), \ P_{out}^{2\times 1} \approx x, \quad (19)$$

so that $P_{out}^{1\times 1} > P_{out}^{1\times 2} \geq P_{out}^{2\times 1}$, and $P_{out}^{1\times 1}/P_{out}^{1\times 2}$ or $P_{out}^{1\times 1}/P_{out}^{2\times 1}$ grow unbounded as $x \to 0$, even though the diversity gain = 1 in all three cases. Note also that $P_{out}^{2\times 1} = P_R \leq P_{out}^{1\times 2}$ in the low outage regime, where $P_R \approx x$ is the outage probability of the Rayleigh $1 \times 1$ channel (no relay), i.e. the $2 \times 1$ relay channel is equivalent to the classical $1 \times 1$ Rayleigh channel (no relay) but the $1 \times 2$ one is not, unless the impact of relay noise is negligible (i.e. $\alpha = K_r G_{rd} \sigma_r^2/\sigma_0^2 \ll 1$). This is the case in (19) when $\alpha \ll \ln\frac{1}{x}$ and $\alpha \ll 1$ for the $1 \times 1$ and $1 \times 2$ channels respectively, and it is always negligible for the $2\times 1$ channel in the low-outage regime (we attribute this to the higher diversity order of the source-relay link in this channel).

As Fig. 3 demonstrates, the approximations in (19) are accurate for $x < 1$ (low outage regime). They also provide an insight into the way typical outage events occur in the relay channels: for $1 \times 2$ and $2 \times 1$ channels, it is when the lowest diversity order link is in outage (S-R and R-D links respectively) and the other link's contribution is negligible. For $1\times 1$ channel, it is not the case anymore since $P_{out}^{1\times 1}$ were about $(2+\alpha)x$ otherwise. Furthermore, $(2+\alpha)x \ll x(\alpha + \ln\frac{1}{x})$ as $x \to 0$, which tells us that the typical outage events are when both links are in partial outage.

[5] It can be shown that the impact of relay noise is negligible so that (17) can be used if $\alpha \ll 1/(m(n-m))$ for $n > m$ (the normalized contribution of the relay noise does not exceed $m(n-m)\alpha$ in this case), and if $\alpha \ll 1$ for $n = m$.

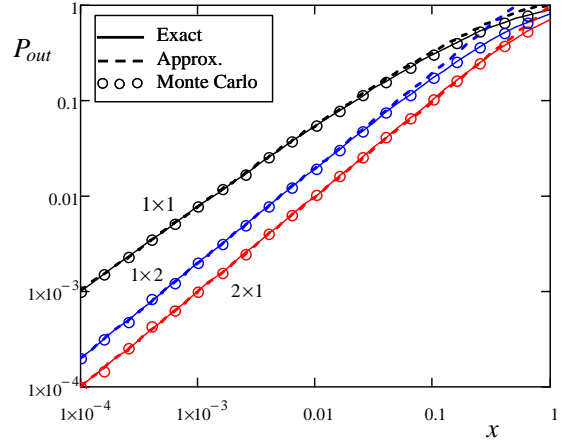

Fig. 3. Outage probability of $1 \times 1$, $2 \times 1$ and $1 \times 2$ relay channels in (18) with corresponding approximations in (19) for $\alpha = 1$. The low-outage approximations are accurate for $x < 1$, and Monte-Carlo simulations validate the analysis.

Let us now compare the $m \times m$ relay channel to the $m \times 1$ MRC channel when $\alpha \to 0$, which has the same diversity order and whose outage probability $P_{MRC} \approx x^m/m!$ at the low outage regime, so that $P_{out}^{m\times m}/P_{MRC} \approx \ln\frac{1}{x}/(m-1)!$, which also grows unbounded at low outage /large SNR regime, even though they have the same diversity order. Thus, the $m\times 1$ MRC channel performs much better than $m \times m$ single-relay channel at low outage. Comparing the MRC channel to $(m+1) \times m$ relay channel, $P_{out}^{(m+1)\times m}/P_{MRC} \approx 1/m! < 1$, i.e. the $(m+1) \times m$ relay channel performs better, which re-enforces our earlier conclusion that one extra source/destination antenna improves the performance significantly, even though it does not affect the diversity order (and also the DMT – see Section IV). Typical outage events can be identified in a similar way.

To explore the impact of correlation, let us now consider the $2 \times 1$ correlated relay channel when $\alpha \to 0$. In this case

$$\mathbf{R}_{rd} = \begin{bmatrix} 1 & \rho \\ \rho^* & 1 \end{bmatrix}$$

where $\rho$ is the (scalar) normalized correlation. The eigenvalues are $\eta_{1,2} = 1 \pm |\rho|$, and the outage probability in (7) can be explicitly expressed as

$$P_{out} = 1 - \frac{1+|\rho|}{2|\rho|}\sqrt{\frac{4x}{1+|\rho|}}K_1\left(\sqrt{\frac{4x}{1+|\rho|}}\right)$$
$$+ \frac{1-|\rho|}{2|\rho|}\sqrt{\frac{4x}{1-|\rho|}}K_1\left(\sqrt{\frac{4x}{1-|\rho|}}\right) \quad (20)$$
$$\approx \frac{x}{2|\rho|}\ln\frac{1+|\rho|}{1-|\rho|}$$

where the approximation holds in the low outage regime. While the effect of correlation is detrimental in general, it is significant only when the correlation is very high – see Fig.

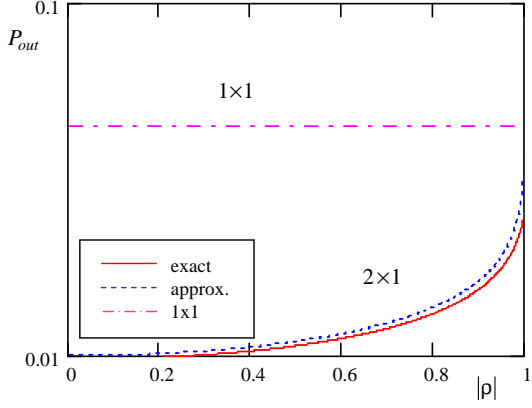

Fig. 4. Outage probability vs. normalized correlation for 2x1 AF single-relay, Rayleigh-fading channel ($\alpha = 0$), at $x = 10^{-2}$ (at $R = 1$bit/s/Hz, this corresponds to $\gamma = 20dB$). Note that correlation has significant effect only when it is very high, $|\rho| > 0.8...0.9$, and that the approximation in (20) is accurate unless $|\rho| \to 1$. Compared to 1x1 channel, the additional destination antenna significantly reduces the outage probability, even for high correlation.

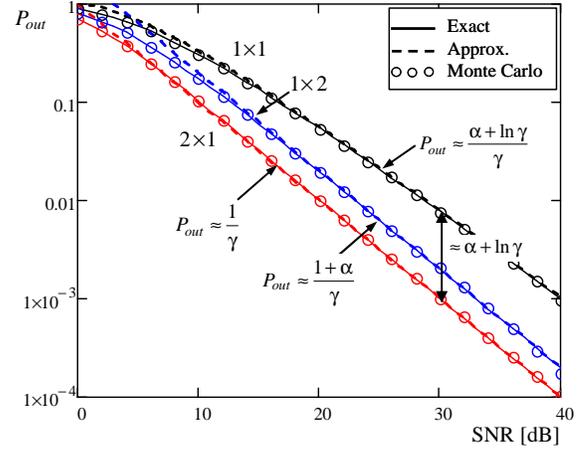

Fig. 5. Outage probability vs. SNR for 1x1, 2x1 and 1x2 AF single-relay i.i.d. Rayleigh-fading channels; $r = 0$, $\alpha = 1$, $10^6$ Monte-Carlo trials. While the SNR-asymptotic DMT is the same for all channels, the gap between the outage probabilities of the 1x1 and 2x1 (1x2) channels increases with SNR; at SNR=40dB, the difference is about 10dB. Unlike an extra Rx antenna, an extra Tx one allows to combat relay noise.

4, and the outage probability increases only logarithmically with correlation. Comparing to the $1 \times 1$ channel, an additional Rx antenna significantly reduces the outage probability, even when the correlation is high. When $\rho \to 0$, (20) reduces to $P_{out}^{2 \times 1}$ in (19), as it should.

### C. Finite-SNR DMT

At finite SNR, the definitions of the diversity and multiplexing gains in [17] are used without the limiting transition (see e.g. [28][29]),

$$d = -\ln P_{out}/\ln \gamma, \quad r = R/\ln \gamma \quad (21)$$

from which it follows that $x = (e^R - 1)/\gamma \approx 1/\gamma^{1-r}$ and the approximation holds at sufficiently high (but finite) SNR, $\gamma^{1-r} \gg 1$. Using this in (14), one obtains for $1 \times 1$, $2 \times 1$ and $1 \times 2$ i.i.d. channels and $0 \leq r < 1$:

$$P_{out}^{1 \times 1} \approx (\alpha + (1-r)\ln \gamma)/\gamma^{1-r},$$
$$P_{out}^{1 \times 2} \approx (1+\alpha)/\gamma^{1-r}, \quad P_{out}^{2 \times 1} \approx 1/\gamma^{1-r} \quad (22)$$

Thus, the diversity gain can be expressed at high SNR/low outage as

$$d_{1 \times 1} \approx 1 - r - \ln(\alpha + (1-r)\ln \gamma)/\ln \gamma$$
$$d_{1 \times 2} \approx 1 - r - \ln(1+\alpha)/\ln \gamma$$
$$d_{2 \times 1} \approx 1 - r$$

It is straightforward to see that the SNR-asymptotic DMT (i.e. when $\gamma \to \infty$) is the same in all three cases: $d(r) = 1 - r$. However, the outage probabilities behave quite differently, as Fig. 5 demonstrates. From (22),

$$\frac{P_{out}^{1 \times 1}}{P_{out}^{2 \times 1}} \approx \alpha + (1-r)\ln \gamma, \quad \frac{P_{out}^{1 \times 2}}{P_{out}^{2 \times 1}} \approx 1 + \alpha, \quad (23)$$

so that $P_{out}^{1 \times 1}/P_{out}^{2 \times 1}$, $P_{out}^{1 \times 1}/P_{out}^{1 \times 2} \to \infty$ as $\gamma \to \infty$ for $r < 1$, i.e. there is a significant advantage in using an extra antenna at either Tx or Rx end at high SNR, even though the SNR-asymptotic DMT is the same in all three cases. Thus, the corresponding DMT-based conclusion breaks down in a most dramatic way. The ultimate reason for this is the $\ln \gamma$ term in (22), not captured by the SNR-asymptotic DMT, which indicates that the latter is ill-suited for relay channels. Furthermore, (23) also shows that an extra Tx rather than Rx antenna is preferable since, unlike the latter, the former eliminates the effect of relay noise at the low outage regime.

When the relay node has full processing capability, i.e. the decode-and-forward protocol with capacity-achieving codes on both ends, the relay channel capacity is $C = \min\{C_{sr}, C_{rd}\}$, where $C_{sr}, C_{rd}$ are the capacities of the source-relay and relay-destination links, so that the weakest link dominates the outage performance, and

$$P_{out}^{1 \times 1} \approx \frac{2}{\gamma^{1-r}}, \quad P_{out}^{1 \times 2} = P_{out}^{2 \times 1} \approx \frac{1}{\gamma^{1-r}} \quad (24)$$

assuming for simplicity the same average SNR on the source-relay and relay-destination links. Clearly, the SNR-asymptotic DMT $d(r) = 1 - r$ is the same in all cases, and also the same as for the AF mode (this is further generalized in Section IV to a broad class of fading distributions). Comparing (24) to (22), we note that the $2 \times 1$ channel in the AF mode outperforms the $1 \times 1$ channel in the DF mode, i.e. an extra Tx antenna in the AF mode is better then the full processing capability at the relay and can be used as a simple alternative of the latter, even though it does not improve the SNR-asymptotic DMT. The same can be said about an extra Rx antenna if $\alpha < 1$. This re-affirms the earlier conclusion that the SNR-asymptotic DMT framework is ill-suited for relay channels and should be used with extreme caution when formulating design guidelines and designing space-time codes. It also follows from (24) that an extra antenna in the DF mode brings a modest SNR gain ($= 3$dB at $r = 0$), unlike the AF mode where this gain can be very significant (see Fig. 5). Comparing (24) to (22), we also note that the full processing capability at the relay does not bring in any advantage for the $2 \times 1$ channel; the same applies to the $1 \times 2$ channel when $\alpha \ll 1$. Clearly, a comparison of different channels/systems in terms of the outage probability may agree with the SNR-asymptotic DMT-based one in some cases while significantly disagreeing in others.

*D. Outage Capacity*

Based on the outage probability results of the previous section, we can now analyse the outage capacity $C_\varepsilon$ which is defined as $C_\varepsilon = \max\{R : P_{out}(R) \leq \varepsilon\}$ and can be expressed as [18]

$$\begin{aligned} C_\varepsilon &= \ln(1 + \gamma x_\varepsilon) \\ C_\varepsilon &\approx \ln \gamma - \ln \frac{1}{x_\varepsilon}, \ \gamma x_\varepsilon \gg 1 \text{ (high SNR)}, \\ C_\varepsilon &\approx \gamma x_\varepsilon, \ \gamma x_\varepsilon \ll 1 \text{ (low SNR)}, \end{aligned} \quad (25)$$

where $x_\varepsilon = P_{out}^{-1}(\varepsilon)$ is the inverse of $P_{out}(x)$, which quantifies the SNR loss compared to the AWGN channel whose capacity is $C_{AWGN} = \ln(1 + \gamma)$. Following (8), the SNR loss can be approximated as

$$\begin{aligned} x_\varepsilon &\approx (\varepsilon/a_1)^{1/m}, \ m < n \\ &\approx (\varepsilon/a_2)^{1/n}, \ m > n \\ &\approx \left( \frac{\varepsilon \cdot m}{a_3 + b_3 \ln \frac{b_3}{\varepsilon}} \right)^{1/m}, \ m = n \end{aligned} \quad (26)$$

i.e. it scales roughly as $\varepsilon^{1/\min(m,n)}$ with the outage probability and increasing $\min(m,n)$ has a significant positive effect on the outage capacity, especially in the small outage region. Since the non-fading AWGN capacity is $C_{AWGN} \approx \ln \gamma$ (high SNR) and $C_{AWGN} \approx \gamma$ (low SNR), the capacity loss in the fading relay link for $m > n$ is

$$\begin{aligned} \Delta C &= C_\varepsilon - C_{AWGN} \approx -\frac{1}{n} \ln \frac{a_2}{\varepsilon} \quad \text{(high SNR)} \\ \Delta &= C_\varepsilon/C_{AWGN} = (\varepsilon/a_2)^{1/n} \quad \text{(low SNR)} \end{aligned} \quad (27)$$

i.e. an additive loss at high SNR and multiplicative at low, so that this effect is much more severe in the latter case. Similar conclusions can also be obtained for the $n > m$ and $n = m$ cases.

To obtain some insight, let us compare $1 \times 1$ and uncorrelated $1 \times 2$, $2 \times 1$ channels

$$\begin{aligned} x_\varepsilon &\approx \frac{\varepsilon}{\alpha + \ln \frac{1}{\varepsilon}}, \ 1 \times 1 \\ &\approx \varepsilon/(1+\alpha), \ 1 \times 2 \\ &\approx \varepsilon, \ 2 \times 1 \end{aligned} \quad (28)$$

The ratio of the SNR loss factors for $2 \times 1 (1 \times 2)$ and $1 \times 1$ channels grows unbounded as $\ln \frac{1}{\varepsilon}$ when $\varepsilon \to 0$, i.e an extra Rx or Tx antenna has a significant positive impact. Fig. 6 compares the normalized outage capacities of $1 \times 1$ and $2 \times 1$ channels, clearly indicating a significant benefit of an extra antenna, especially in the low SNR regime.

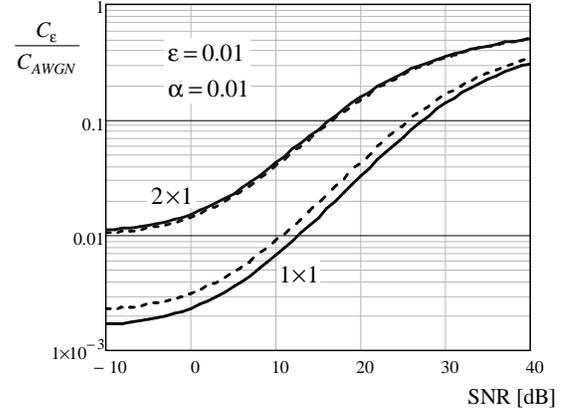

Fig. 6. The outage capacity of 1x1 and 2x1 channels normalized to the AWGN channel capacity versus SNR; solid line – Monte-Carlo simulations, dashed line – via the approximation in (26) and (25). While an extra Tx antenna brings 70% increase in the outage capacity at high SNR = 40 dB, it brings much more dramatic, 5-fold increase at low SNR = -10 dB.

## IV. SNR-ASYMPTOTIC DMT

Following [17][18][21], we define the SNR-asymptotic diversity and multiplexing gains as

$$d = -\lim_{\gamma \to \infty} \ln P_{out}/\ln \gamma, \quad r = \lim_{\gamma \to \infty} R/\ln \gamma \quad (29)$$

Using the results of Section III, the SNR-asymptotic diversity-multiplexing tradeoff in the single-relay correlated channel can now be characterised.

*Theorem 3:* Consider the relay channel in (1) under the conditions of Theorem 1 (correlated non-identically distributed fading) or Theorem 2 (i.i.d. fading). Its SNR-asymptotic DMT can be expressed as

$$d(r) = \min(m, n)(1 - r), \quad 0 \leq r \leq 1 \quad (30)$$

*Proof:* it follows from (29) that $R = r \ln \gamma$ and $x = (e^R - 1)/\gamma = 1/\gamma^{1-r}$ as $\gamma \to \infty$. Using this in (8) to obtain $P_{out}$ and substituting it in (29), one obtains $d$ as in (30). In the case of i.i.d. fading, (14) is used instead of (8) and the same result is obtained. ∎

Theorem 1 in [21] (for single relay case) is a special case of Theorem 3 here, i.e. when $\mathbf{h}_{sr}$ and $\mathbf{h}_{rd}$ are i.i.d. complex Gaussian. The DMT of relay channels/networks under i.i.d. Rayleigh/Rician fading has been also studied in [34][35]. Following a different line of analysis [30][31], a significant generalization of this result can be obtained for a broad class of fading distributions (including, as special cases, Rayleigh, Rice, Nakagami, and Weibull, which may be correlated, non-zero mean and non-identically distributed[6]).

*Theorem 4:* Consider the relay channel in (1) such that the PDFs $f_s(x)$ and $f_d(x)$ of $|\mathbf{h}_{sr}|^2$ and $|\mathbf{h}_{rd}|^2$ behave polynomially near zero, i.e. $f_s(x) \sim x^{d_s - 1}$, $f_d(x) \sim x^{d_d - 1}$ as $x \to 0$, where $d_s, d_d$ are the diversity gains (orders) of the source-relay and relay-destination links at $r = 0$, and $f(x) \sim g(x)$ means

---

[6] It is straightforward to show that full-rank correlation does not affect the order of the polynomial behavior of $|\mathbf{h}_{sr}|^2$ or $|\mathbf{h}_{rd}|^2$ near zero.

8that there exist positive constants $0 < A \leq B < \infty$, such that $Ag(x) \leq f(x) \leq Bg(x)$ for sufficiently small $x$. The DMT of this channel in the amplify-and-forward mode is

$$d(r) = \min(d_s(r), d_d(r)) \qquad (31)$$
$$= \min(d_s, d_d)(1-r), \quad 0 \leq r \leq 1$$

where $d_s(r) = d_s(1-r), d_d(r) = d_d(1-r)$ are the SNR-asymptotic DMTs of the source-relay and relay-destination links. ∎

Theorem 4 demonstrates that the SNR-asymptotic DMT is affected by the number of degrees of freedom (diversity order) available in the channel and not by particular fading distribution, as long as the definition of diversity gain in (29) makes sense. Similar result has been also established for full-rank MIMO channels in [22]. It follows from Theorem 4 that full-rank Tx/Rx correlation and the relay noise do not affect the SNR-asymptotic DMT[7]. Furthermore, from (4) and (5), the absence of CSI at the source is equivalent to an $m$-fold SNR loss and, therefore, has no effect on the SNR-asymptotic DMT (i.e. (31) still holds). The transmit beamforming in combination with QAM modulation (e.g. see [18]) is an example of a space-time code that achieves the DMT of the single-relay channel with the source CSI. When no such CSI is available, isotropic transmission in combination with QAM will achieve the DMT.

Let us now consider the DMT of the decode-and-forward single-relay channel, assuming capacity achieving codes and complete decoding/encoding at the relay. Following similar line of analysis, one can establish its DMT for the broad class of fading distributions.

*Theorem 5:* Under the conditions of Theorem 4, the diversity-multiplexing tradeoff of the single-relay channel in the DF mode is the same as in the AF one and is given by (31), i.e. full processing capability at the relay does not improve the DMT. ∎

Thus, the single-relay channel subject to fading from a broad class of distributions has the same SNR-asymptotic DMT in the amplify-and-forward and decode-and-forward modes. However, as we argued in section III, the full processing capability at the relay does help to reduce significantly the outage probability when the SNR is finite. Furthermore, following Theorem 3, the DMT is affected by $\min(m,n)$[8], so that from the SNR-asymptotic DMT viewpoint there is no point in using unequal number of antennas (e.g. no point to use more source than destination antennas). This, however, does not hold when SNR is finite as our analysis in Section III shows: an additional source or destination antenna can be traded-off for the full processing capability at the relay and improves the performance significantly. Therefore, the SNR-asymptotic DMT should be used with caution when formulating design guidelines. This conclusion also applies to full-rank MIMO channels for a broad class of fading distributions [29].

---

[7] This has been established for Rayleigh-fading keyhole channels in [24].

[8] In the case of i.i.d. Rayleigh and Rician-fading links, this conclusion was obtained in [21] and [35].

## V. SELECTION RELAYING

Let us now consider the case of multiple relay nodes and when the selection relaying is used, i.e. only the best relay node is used at any time, out of $N$ available nodes. This protocol was originally proposed in [16] and was also considered in [14] in the low-SNR regime. It is motivated by its low complexity and also by the fact that little interference is created to other users since only one relay is transmitting. We assume that different relay links are independent of each other (which is justified by geographical separation of the relays), that each link follows the statistics of section III (for outage probability analysis), and that amplify-and-forward protocol is used[9] (unless indicated otherwise).

### A. Outage Probability

An outage takes place when the best relay link and therefore all the relay links are in outage, so that the outage probability is

$$P_{out} = \Pr\{C < R\} = \prod_{i=1}^{N} P_i \qquad (32)$$

where $P_i = \Pr\{C_i < R\}$ is the outage probability of $i$-th relay link, which is given in (7)-(20), and each link is allowed to have its own statistics. When all the links have the same statistics, (32) reduces to

$$P_{out} = P_1^N \qquad (33)$$

so that the outage probability decreases exponentially in $N$ compared to the single relay case, which is especially significant in the low outage regime, when $P_1 \ll 1$. From (33) and (8), (14), the diversity order is $d = N\min(m,n)$ under Rayleigh fading (also with full-rank correlation), so that the number of antennas and relay nodes can be traded off for each other.

To obtain further insight, let us consider $1 \times 1$, $1 \times 2$ and $2 \times 1$ i.i.d. channels, for which (33) simplifies in the low outage regime to

$$P_{out}^{1\times 1} \approx x^N \left(\alpha + \ln \frac{1}{x}\right)^N, \qquad (34)$$
$$P_{out}^{1\times 2} \approx x^N (1+\alpha)^N, \ P_{out}^{2\times 1} \approx x^N,$$

so that the ratio $P_{out}^{1\times 1}/P_{out}^{2\times 1} \approx \left(\alpha + \ln \frac{1}{x}\right)^N$ grows unbounded as $x \to 0$ (i.e. $\gamma \to \infty$ under fixed $R$ or $R \to 0$ under fixed $\gamma$) and, as in the case of $N = 1$, there is a significant advantage in using 2 source antennas instead of 1 at this regime, even though the diversity gain is the same in both cases. This conclusion also holds when the source and destination are equipped with more antennas.

### B. Diversity-Multiplexing Tradeoff

Based on the outage probability in (32) and Theorems 3, 4, the SNR-asymptotic DMT of selection relaying can be immediately characterized for a broad class of fading distributions.

---

[9] In this case, the best relay is $\arg\max_i \left\{ |\mathbf{h}_{rd,i}|^2 |\mathbf{h}_{sr,i}|^2 / (1 + \alpha |\mathbf{h}_{rd,i}|^2) \right\}$.

*Theorem 6:* Consider the relay channel in (1) under the conditions of Theorem 4 and 5 and assume that all relay links are independent of each other. Its diversity-multiplexing tradeoff in the AF and DF modes is as follows:

$$d(r) = \sum_{i=1}^{N} \min(d_{s,i}(r), d_{d,i}(r)) \quad (35)$$
$$= (1-r) \sum_{i=1}^{N} \min(d_{s,i}, d_{d,i}), \quad 0 \leq r \leq 1,$$

where $d_{s,i}(r) = d_{s,i}(1-r), d_{d,i}(r) = d_{d,i}(1-r)$ are the DMTs of the Tx-relay and relay-Rx links. ∎

It follows that the total DMT is the sum of the DMTs for each relay and increasing the maximum of $(d_{s,i}, d_{d,i})$ (by increasing the number of antennas at the corresponding end of the link) will not improve the DMT if $\min(d_{s,i}, d_{d,i})$ are fixed, which is the same as for the single relay case.

In the case of identical link statistics, (35) simplifies to

$$d(r) = N(1-r)\min(d_s, d_d), \quad 0 \leq r \leq 1 \quad (36)$$

and $N$-fold increase in the diversity gain is obvious. However, we caution that the same limitations of DMT-based analysis/design as for the single-relay case also hold for selection relaying.

Finally, we comment that the impact of direct link (source-destination) can also be included in the analysis in the same way: any reasonable relaying protocol makes use of both links so that an outage takes place when both links are in outage and the overall outage probability is $P_{out}P_{dl}$, where $P_{dl}$ is the outage probability of the direct link (this is exactly the case when selection relaying is used), and the overall diversity gain is the sum of per-link diversity gains.

## VI. CONCLUSION

Outage probability and diversity-multiplexing tradeoff have been investigated for MIMO relay channels. The SNR-asymptotic DMT has been established for such channels under a broad class of fading distributions, thus generalizing earlier results known for i.i.d. and correlated Rayleigh channels. The latter two have been investigated in greater depth. Compact, closed-form expressions, and corresponding low-outage approximations have been obtained for their outage probability and capacity, which, unlike the SNR-asymptotic DMT, also hold at realistic SNR values. Significant difference between the SNR-asymptotic DMT and finite-SNR outage performance has been emphasized. In particular, while the SNR-asymptotic DMT is not improved by using more antennas on either side, the outage probability can be significantly improved and, in particular, an extra Tx antenna can be traded-off for full processing capability at the relay. The results are extended to channels with multiple relays under selection relaying. The SNR loss of fading relay channels compared to the AWGN channel and the impact of relay noise have been quantified. Under certain conditions, the relay channel has been shown to be equivalent to the maximum ratio combining channel.

## APPENDIX

### A. Proof of Theorem 1

Let $g_s = |\mathbf{h}_{sr}|^2$, $g_d = |\mathbf{h}_{rd}|^2$. Since they are independent and non-negative,

$$P_{out} = \Pr\left\{\frac{g_s g_d}{1+\alpha g_d} < x\right\} \quad (37)$$
$$= \int_0^\infty f_d(t_1) \int_0^{x(1+\alpha t_1)/t_1} f_s(t_2) dt_2 dt_1,$$

where $f_s(t)$ and $f_d(t)$ are PDFs of $g_s$ and $g_d$. Under the adopted assumptions, the distribution of $g$ is generalized $\chi^2$ with the characteristic function $\Phi_g(\omega) = \det[\mathbf{I} - j\omega\mathbf{R}]^{-1}$, where $\mathbf{R}$ is the correlation matrix and $j = \sqrt{-1}$. When $\mathbf{R}$ is non-singular and has $N$ distinct eigenvalues $\lambda_k$, the characteristic function (CF) of $g$ is

$$\Phi_g(\omega) = \prod_{k=1}^N (1-j\omega\lambda_k)^{-1} = \sum_{k=1}^N A_k (1-j\omega\lambda_k)^{-1}, \quad (38)$$

where $A_k$ are the coefficients of the partial fraction decomposition of $\Phi_g(\omega)$. Thus, the PDF of $g$ is:

$$f_g(x) = \frac{1}{2\pi}\int_{-\infty}^\infty \Phi_g(\omega)e^{-j\omega x}d\omega = \sum_{k=1}^N \frac{A_k}{\lambda_k} e^{-x/\lambda_k}, \quad (39)$$

where $x \geq 0$. After substituting (39) into (37) and integrating using the standard integrals [27], (7) follows. ∎

### B. Proof of Corollary 1.1

Using the series expansion of $K_N(x)$ [27],

$$K_N(x) = \frac{1}{2}\sum_{k=0}^{N-1}(-1)^k \frac{(N-k-1)!}{k!(x/2)^{N-2k}} \quad (40)$$
$$+(-1)^{N+1}\sum_{k=0}^\infty \frac{(x/2)^{N+2k}}{k!(N+k)!}$$
$$\times \left(\ln\frac{x}{2} - \frac{1}{2}\psi(k+1) - \frac{1}{2}\psi(N+k+1)\right)$$

and making use of the following properties of partial fraction decomposition,

$$\sum_{k=1}^N A_k = 1, \quad \sum_{k=1}^N \frac{A_k}{\lambda_k^i} = 0, \; i=1,...,N-1, \quad (41)$$

one obtains after lengthy but straightforward manipulations an alternative expression for (7):

(i) when $m \geq n$,

$$P_{out} = \sum_{k=m}^\infty \sum_{j=1}^m \frac{(-1)^{k+1}\alpha^k}{k!}\frac{A_j}{\lambda_j^k}x^k \quad (42)$$
$$+\sum_{k=n}^\infty \sum_{l=k}^\infty \sum_{i,j} \frac{A_j B_i}{\lambda_j^l \eta_i^k} D_{kl}(\alpha) x^l \left(\ln\frac{\lambda_j}{x} + \Psi_k\right)$$
$$+\sum_{k=1}^\infty \sum_{l=\max(k,m)}^\infty \sum_{i,j} \frac{A_j B_i \ln\eta_i}{\lambda_j^l \eta_i^k} D_{kl}(\alpha) x^l$$

(ii) when $m < n$,

$$P_{out} = \sum_{k=m}^\infty \sum_{j=1}^m \frac{(-1)^{k+1}\alpha^k}{k!}\frac{A_j}{\lambda_j^k}x^k \quad (43)$$
$$+\sum_{k=1}^\infty \sum_{l=\max(k,m)}^\infty \sum_{i,j} \frac{A_j B_i}{\lambda_j^l \eta_i^k} D_{kl}(\alpha) x^l \left(\ln\frac{\eta_i}{x} + \Psi_k\right)$$
$$+\sum_{k=n}^\infty \sum_{l=k}^\infty \sum_{i,j} \frac{A_j B_i \ln\lambda_j}{\lambda_j^l \eta_i^k} D_{kl}(\alpha) x^l;$$

The leading term in the above series corresponds to $k = m$ when $m \leq n$, and $k = n$ when $m > n$. Keeping only this term and making use of the following property of partial fraction decomposition:

$$\sum_{k=1}^N \frac{A_k}{\lambda_k^N} = (-1)^{N-1}\prod_k \frac{1}{\lambda_k} = (-1)^{N-1}\det\mathbf{R}^{-1} \quad (44)$$

one obtains (8) after some manipulations. ∎

### C. Proof of Theorem 2

Under the adopted assumptions, $g$ is the central $\chi^2$ random variable with the following PDF: $f_g(x) = x^{N-1}e^{-x}/(N-1)!$, where $N$ ($m$ or $n$) is the number of degrees of freedom. After substituting this into (37), integrating (using standard integrals in [27]), and making some manipulations, (12) follows. ∎

$$f_l(\alpha) = \sum_{i=0}^{m}\sum_{j=[l+1-p_i]_+}^{l} \Omega_{ij}(\alpha)\frac{(-1)^l(p_0-q_{lij}-1)!}{q_{l0j}!(l+m)} - c_l(\alpha),\ m<n, \qquad (45)$$

$$= \sum_{i=1}^{m}\sum_{j=[l+1-p_i]_+}^{l} \Omega_{ij}(\alpha)\frac{(-1)^l(-q_{lij}-1)!}{q_{l0j}!(l+m)} - c_l(\alpha),\ m=n;$$

$$g_l(\alpha) = \sum_{i=0}^{\min(l-p_0,m)}\sum_{j=0}^{l-p_0-i} \Omega_{ij}(\alpha)\frac{(-1)^{p_i+j+1}}{(q_{lij}-p_0)!q_{l0j}!(l+m)},\ l\geq p_0,\ \text{and 0 otherwise};$$

$$c_l(\alpha) = \sum_{i=0}^{\min(l-p_0,m)}\sum_{j=0}^{l-p_0-i} (-1)^{p_i+j+1}\Omega_{ij}(\alpha)\frac{\psi(q_{lij}-p_0+1)+\psi(q_{l0j}+1)+(l+m)^{-1}}{(q_{lij}-p_0)!q_{l0j}!(l+m)},\ l\geq p_0,\ \text{and 0 otherwise};$$

$$\Omega_{ij}(\alpha) = \frac{\alpha^{i+j}m}{i!j!(n-1)!(m-i)!},\quad p_i=|m-n-i|,\quad q_{lij}=l-i-j;$$

$$f'_l(\alpha) = \sum_{i=0}^{\min(l,p_0-1)}\sum_{j=[l+1-p_0]_+}^{l-i} \Omega_{ij}(\alpha)\frac{(-1)^{l-i}(p_0-q_{l0j}-1)!}{q_{lij}!(l+n)}, \qquad (46)$$

$$f''_l(\alpha) = \sum_{i=p_0+1}^{m}\sum_{j=[l+1-i]_+}^{l-p_0} \Omega_{ij}(\alpha)\frac{(-1)^{l-p_0}(-q_{lij}-1)!}{(q_{l0j}-p_0)!(l+n)},\ l\geq p_0,\ \text{and 0 otherwise};$$

$$c'_l(\alpha) = \sum_{i=p_0+1}^{\min(l,m)}\sum_{j=0}^{l-i} \Upsilon_{ijl}\Omega_{ij}(\alpha),\quad g'_l(\alpha) = \sum_{i=p_0+1}^{\min(l,m)}\sum_{j=0}^{l-i} \Theta_{ijl}\Omega_{ij}(\alpha),\ l\geq p_0+1,\ \text{and 0 otherwise};$$

$$c''_l(\alpha) = \sum_{i=0}^{p_0}\sum_{j=0}^{l-p_0} \Upsilon_{ijl}\Omega_{ij}(\alpha),\quad g''_l(\alpha) = \sum_{i=0}^{p_0}\sum_{j=0}^{l-p_0} \Theta_{ijl}\Omega_{ij}(\alpha),\ l\geq p_0,\ \text{and 0 otherwise};$$

$$\Theta_{ijl} = \frac{(-1)^{p_i+j+1}}{q_{lij}!(q_{l0j}-p_0)!(l+n)},\quad \Upsilon_{ijl} = \left(\psi(q_{lij}+1)+\psi(q_{l0j}-p_0+1)+(l+n)^{-1}\right)\Theta_{ijl}$$

### D. Proof of Corollary 2.1

Using the series expansion of $K_N(x)$ (see (40)) and (12), one obtains after lengthy but otherwise straightforward manipulations, (13), where $f_l(\alpha)$ and $g_l(\alpha)$ are independent of $x$. When $m \leq n$ $f_l(\alpha)$ and $g_l(\alpha)$ are given by (45), where $[x]_+ = x$ if $x \geq 0$ and 0 otherwise. When $m > n$,

$$f_l(\alpha) = f'_l(\alpha) + f''_l(\alpha) - c'_l(\alpha) - c''_l(\alpha),$$
$$g_l(\alpha) = g'_l(\alpha) + g''_l(\alpha),$$

where $f'_l(\alpha)$, $f''_l(\alpha)$, $g'_l(\alpha)$, $g''_l(\alpha)$, $c'_l(\alpha)$ and $c''_l(\alpha)$ are independent of $x$, and given by (46). The first non-zero term in the above series corresponds to $l=0$, so that the lowest power of $x$ is $\min(m,n)$. By keeping this term, one obtains (14). ∎

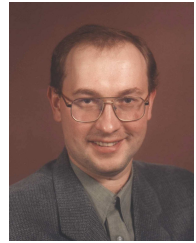

**Sergey Loyka** (M'96–SM'04) was born in Minsk, Belarus. He received the Ph.D. degree in Radio Engineering from the Belorussian State University of Informatics and Radioelectronics (BSUIR), Minsk, Belarus in 1995 and the M.S. degree with honors from Minsk Radioengineering Institute, Minsk, Belarus in 1992. Since 2001 he has been a faculty member at the School of Information Technology and Engineering, University of Ottawa, Canada. Prior to that, he was a research fellow in the Laboratory of Communications and Integrated Microelectronics (LACIME) of Ecole de Technologie Superieure, Montreal, Canada; a senior scientist at the Electromagnetic Compatibility Laboratory of BSUIR, Belarus; an invited scientist at the Laboratory of Electromagnetism and Acoustic (LEMA), Swiss Federal Institute of Technology, Lausanne, Switzerland. His research areas include wireless communications and networks, MIMO systems and smart antennas, RF system modeling and simulation, and electromagnetic compatibility, in which he has published extensively. He received a number of awards from the URSI, the IEEE, the Swiss, Belarus and former USSR governments, and the Soros Foundation.

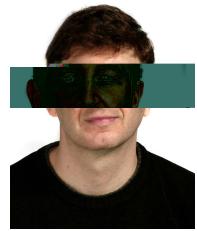

**Georgy Levin** received the B.S. and M.S. degrees, both cum laude, in Electrical and Computer Engineering from Ben-Gurion University of the Negev, Israel in 1995 and 2000, and the Ph.D. degree from the University of Ottawa, Ontario, Canada in 2008. He is currently a research assistant at the University of Ottawa. His research spans the fields of wireless communications and information theory with specific interest in MIMO systems, smart antennas, relay networks and cognitive radio. He received a number of awards from Canada, Israel and former USSR governments.